\documentclass{aa}

\usepackage{graphicx}
\usepackage{txfonts}
\usepackage[flushleft]{threeparttable} 
\usepackage{hyperref}
\hypersetup{colorlinks=true,linkcolor=blue,citecolor=blue,filecolor=blue,urlcolor=blue}

\usepackage{xcolor}

\newcommand{\ha}{H$\alpha$}

\newcommand{\hei}{He~{\sc i}}

\newcommand{\nh}{$N_\mathrm{H}$}

\def\kms{\mbox{${\rm km}\:{\rm s}^{-1}\:$}}

\def\lesssim{\mathrel{\hbox{\rlap{\hbox{\lower4pt\hbox{$\sim$}}}\hbox{$<$}}}}

\def\gtrsim{\mathrel{\hbox{\rlap{\hbox{\lower4pt\hbox{$\sim$}}}\hbox{$>$}}}}

\begin{document}

\title{Simultaneous X-ray and optical spectroscopy of V404~Cygni supports the multi-phase nature of X-ray binary accretion disc winds}

   \author{Teo Muñoz-Darias
          \inst{\ref{i1},\ref{i2}}
          \and
          Gabriele Ponti\inst{\ref{i3},\ref{i4}}
          }

   \institute{Instituto de Astrofísica de Canarias, E-38205 La Laguna, Tenerife, Spain \label{i1}
         \and
             Departamento de Astrofísica, Universidad de La Laguna, E-38206 La Laguna, Tenerife, Spain \label{i2}
        \and INAF-Osservatorio Astronomico di Brera, Via E. Bianchi 46, I-23807 Merate (LC), Italy \label{i3}
        \and Max-Planck-Institut f{\"u}r extraterrestrische Physik, Giessenbachstrasse, D-85748, Garching, Germany \label{i4}
             }

\titlerunning{A multi-phase disc wind in V404~Cygni}
\authorrunning{Muñoz-Darias \& Ponti}
 
  \abstract{Observational signatures of accretion disc winds have been found in a significant number of low-mass X-ray binaries at either X-ray or optical wavelengths. The 2015 outburst of the black hole transient V404 Cygni provided a unique opportunity for studying both types of outflows in the same system. We used contemporaneous X-ray (\textit{Chandra} Observatory) and optical (\textit{Gran Telescopio Canarias}, GTC) spectroscopy, in addition to hard X-ray light curves (\textit{INTEGRAL}). We show that the kinetic properties of the wind, as derived from P-Cyg profiles detected in the optical range at low hard X-ray fluxes and in a number of X-ray transitions during luminous flares, are remarkably similar. Furthermore, strictly simultaneous data taken at intermediate hard X-ray fluxes show consistent emission line properties between the optical and the X-ray emission lines, which most likely arise in the same accretion disc wind. We discuss several scenarios to explain the properties of the wind, favouring  the presence of a dynamic, multi-phase outflow during the entire outburst of the system. This study, together with the growing number of wind detections with fairly similar characteristic velocities at different wavelengths, suggest that wind-type X-ray binary outflows might be predominantly multi-phase in nature.}


   \maketitle


\section{Introduction}
\defcitealias{King2015}{K15} 
\defcitealias{Munoz-Darias2016}{MD16} 
\defcitealias{MataSanchez2018}{MS18}



Active low-mass X-ray binaries (LMXBs), where a stellar-mass black hole or a neutron star is accreting material from a low-mass donor, provide one of the best opportunities for studying accretion and outflows processes in strong gravity conditions (e.g. \citealt{Fender2004}). Multiwavelength observations of these systems (e.g. \citealt{Corral-Santana2016, Tetarenko2016}) have revealed the presence of two main (X-ray) accretion states (hard and soft),  as well as associated outflows in the form of collimated jets (mainly detected in radio waves) and massive winds (e.g. \citealt{Remillard2006b, Done2007, Belloni2011, Fender2016} for reviews).  In particular, accretion disc winds have been detected in a number of LMXBs at (mainly) X-ray and optical wavelengths. They are thought to have a fundamental impact on the accretion process, as they can carry significant amounts of mass and angular momentum (see e.g. \citealt{Neilsen2011, Ponti2012, Tetarenko2018, Casares2019, Dubus2019}). 

The so-called hot winds are detected via X-ray spectroscopy in systems seen at high orbital inclination (i.e. closer to edge-on) during relatively soft accretion states (e.g. \citealt{Ueda1998, Miller2006, Neilsen2009, Ponti2012, Ponti2014, DiazTrigo2016}). Likewise, optical spectroscopy has revealed the presence of P-Cyg line profiles and other conspicuous signatures of colder winds in several LMXBs (e.g. \citealt{Munoz-Darias2016}, hereafter \citetalias{Munoz-Darias2016}; \citealt{Munoz-Darias2018}). More recently, the intense follow-up of the black-hole transient MAXI ~J1820+070 throughout its entire discovery outburst showed that optical P-Cyg profiles were solely present during the hard state, while near-infrared wind signatures could be still detected during the soft state (\citealt{Munoz-Darias2019, SanchezSierras2020}). This picture is consistent with optical and near-infrared observations of up to ten other systems (see Table 2 in \citealt{PanizoEspinar2022}), including outbursts with more limited coverage (e.g. \citealt{MataSanchez2022}), hard state-only outburst (e.g. \citealt{Cuneo2020}), and persistent sources (\citealt{Bandyopadhyay1997}).  While the above works have demonstrated that both hot and cold winds are common (if not ubiquitous) in X-ray binaries, it is unclear whether they are different types of outflows or whether they represent different faces of the same phenomenon. Simultaneous wind detections have been reported in the optical and the near-infrared (\citealt{SanchezSierras2020}), as well as in the optical and the ultraviolet (\citealt{Munoz-Darias2020,CastroSegura2022}) regimes. However, a simultaneous study of hot (X-ray) and cold (optical) wind-type outflows has not been possible to date. 

The 2015 outburst of the black hole transient V404 Cygni was arguably among the most revealing ones ever recorded.  During this luminous and violently variable event, the system displayed a rather unique phenomenology across the electromagnetic and time domains  (e.g. \citealt{Rodriguez2015, Kimura2016, Gandhi2016, Siegert2016, Motta2017a, Tetarenko2017, MillerJones2019}). However, the outburst was surprisingly short, with the most active phase merely lasting two weeks, which has been proposed to be a consequence of the massive wind that was repeatedly detected via optical spectroscopy (\citetalias{Munoz-Darias2016}; \citealt{MataSanchez2018}, hereafter \citetalias{MataSanchez2018}; \citealt{Casares2019}). Likewise, conspicuous features indicating the presence of a X-ray wind were revealed by two Chandra pointings (\citealt{King2015}, hereafter \citetalias{King2015}). In this work, we present a detailed analysis of the simultaneous X-ray and optical observational signatures of the wind-type outflow detected during this event.



\section{Observations and data reduction}
We analysed the two Chandra observations (hereafter, obs-1 and obs-2) taken during the luminous phase of the outburst (see \citealt{Plotkin2017} for additional pointings taken during the decay), together with two contemporaneous spectroscopic epochs taken with the Gran Telescopio Canarias (GTC): one taken 28 minutes after the first Chandra observation ended (hereafter, GTC-1) and another one that was strictly simultaneous with the second \textit{Chandra} window (hereafter, GTC-2). The start and end times for each observation are summarised in Table \ref{log}.   

\begin{table*}[ht!]
        \centering
        \caption{Observing epochs included in the analysis}
        
        \begin{threeparttable}
        \begin{tabular}{l c c c c c c}
        
                \hline
                Observation (2015) & Observatory & Start (UT) & End (UT) & Duration (ks) & Comments
                \\
                \hline
                \hline
                Obs-1 (June 22)  & \textit{Chandra} & 13:40 & 22:41& 21 & \\
                GTC-1(June 22--23; day-6 in \citetalias{Munoz-Darias2016}) & GTC & 23:09 & 01:06 & 7 &  \\
                Obs-2 (June 23--24)  & \textit{Chandra} & 21:26 & 05:21& 25 & Fully covered by \textit{INTEGRAL}\\
                GTC-2 (June 24; day-7 in \citetalias{Munoz-Darias2016})& GTC & 01:18 & 02:07 & 2.3 & Simultaneous with obs-2\\
                
                \hline
        \end{tabular}
        \label{log}
        
        \end{threeparttable}
\end{table*} 
\begin{figure*}
    \centering
    \includegraphics[width=18truecm]{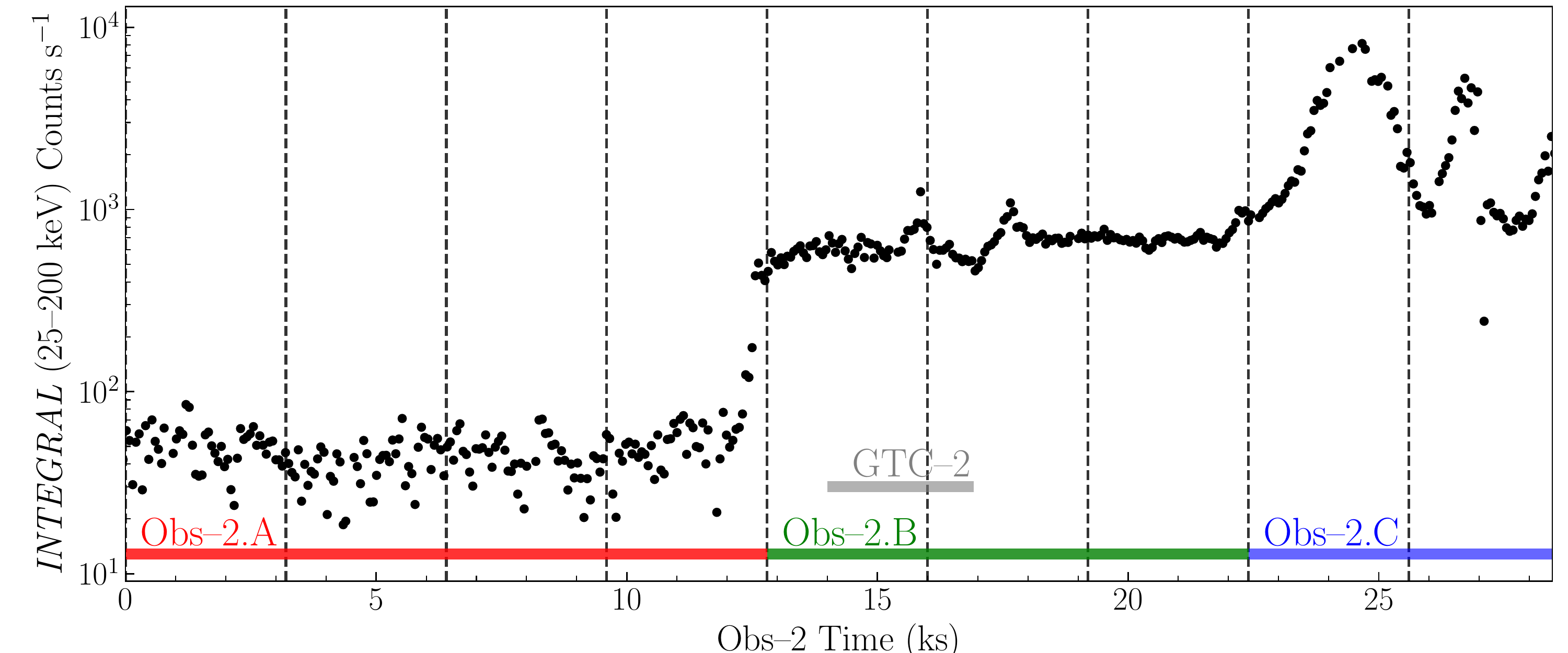}
    \vspace{0.5cm}
    
     \includegraphics[width=18truecm]{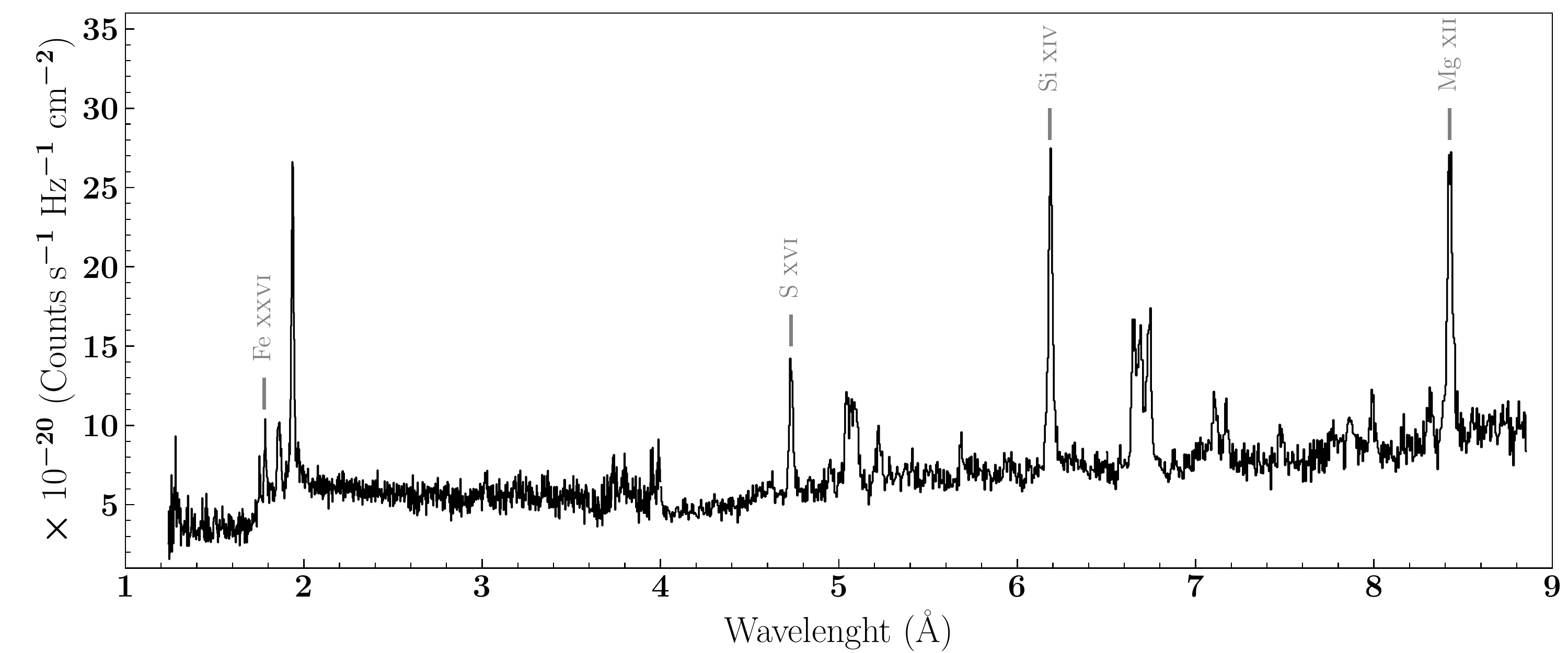}
    \caption{Hard X-ray light curve and Chandra spectrum during obs-2. Top panel: \textit{INTEGRAL} light curve. The dashed vertical lines indicates the boundaries of the 3.2 ks spectra (also used in \citetalias{King2015}). Following the hard X-ray evolution, we have divided the observation into three intervals: obs-2.A (spectra 1 to 4; low flux), obs-2.B (5 to 7; plateau), and obs-2.C (8 to 9; flares). Bottom panel: Combined MEG+HEG spectrum. We have labelled the emission lines included in this analysis (see Section \ref{analysis}).}
    \label{fig:data}%
\end{figure*}

\subsection{X-rays}

In an effort to avoid radiation damage as a result of the exceptional brightness of V404~Cygni, the \textit{Chandra} observations were performed with the ACIS-S array in continuous clocking mode and with the zeroth-order position located off the detector array. 
Therefore, only positive and negative orders are available for the medium-energy grating (MEG) and high-energy grating (HEG), respectively. 
The \textit{Chandra} data were reprocessed with the {\sc chandra\_repro} task of the {\sc ciao} software and the keyword {\sc recreate\_tg\_mask} set as yes, in order to improve the location of the zeroth-order. 
Spectra and response matrices were computed with {\sc tgextract} and {\sc mktgresp} tasks, respectively, applying the proper selection in time. 
To check the accuracy of the zeroth-order location, we compared our spectra with the ones downloaded from the \textit{Chandra} Grating-Data Archive and Catalog (TGCat; \citealt{Huenemoerder2011}\footnote{\url{http://tgcat.mit.edu/}}), finding a good agreement. 

In order to optimise the signal-to-noise ratio (S/N), we used MEG data (with a resolution of $\Delta \lambda \sim 0.023$ \AA) above 6.5 \AA\ ($<$1.9 keV), while the spectral range below 4 \AA\ (i.e. $>3.1$ keV) is covered by HEG data only (at $\Delta \lambda \sim 0.012$ \AA)\footnote{\url{https://cxc.harvard.edu/proposer/POG/html/chap8.html}}. Within 4--6.5 \AA\ we combined both HEG and MEG datasets (at the MEG resolution and wavelength scale). 
In addition, we made use of the X-ray light curve (25--200 keV) from \textit{INTEGRAL} \citep{Winkler2003}, which covers obs-2 entirely.  We used this dataset \citep{Kuulkers2015} with a time resolution of $\sim 64$s (see \citetalias{Munoz-Darias2016} for details and \citealt{Rodriguez2015} for the full \textit{INTEGRAL} analysis during the first part of the outburst). 

\subsection{Optical}
Optical spectroscopy (3630--7500 \AA) was obtained using the instrument OSIRIS \citep{Cepa2000} at the GTC (La Palma, Spain). For this work we used the average spectra for each of the two epochs (see Table \ref{log}), that were already presented in \citetalias{Munoz-Darias2016} and \citetalias{MataSanchez2018}, where we refer the reader for further details on the data reduction. We note that the velocity resolution of this dataset is $\sim 250$ \kms, significantly better than that of MEG and HEG data ($\sim 1100$ and $580$ \kms, respectively at  6 \AA). We focus the study on the \hei-5876 \AA\ (hereafter \hei) and \ha\ emission lines, the two optical transitions showing the strongest wind features throughout the outburst. These recombination lines are arguably the most relevant optical wind tracers in LMXBs, accreting white dwarfs (e.g. \citealt{Kafka2004}) and massive stars (e.g. \citealt{Prinja1996}).   



\section{Analysis and results}
\label{analysis}
An analysis of the \textit{Chandra} data was presented in \citetalias{King2015}. To allow for an easy comparison with this work, we initially divided each \textit{Chandra} pointing into $10\times3.2$ ks  (hereafter, spec-1.1 to -1.10)  and $9\times3.2$ ks (hereafter, spec-2.1 to -2.9) spectra, respectively. 
As reported in \citetalias{King2015}, the data are very rich in emission lines, which show significant variability across the $19\times3.2$~ks intervals. Figure \ref{fig:data}~(bottom panel) shows the average spectrum of obs-2, which is qualitatively very similar to that of obs-1. This work is focussed on the analysis of the recombination line of Si~\textsc{xiv} (6.184 \AA; 2.005 keV). This is the strongest spectral feature in the soft X-ray band and is covered by both MEG and HEG data. Furthermore, it is a singlet (Ly$\alpha$) and is located in a spectral region with a relatively flat continuum. These properties make it particularly suited for identifying blue-shifted absorptions and their associated blue-edge velocities. In addition, we also included  S~\textsc{xvi} (4.732 \AA; 2.620 keV) and Mg~\textsc{xii} (8.423 \AA; 1.472 keV), which are also Ly$\alpha$ transitions sharing to some extent some of the above properties.  Each of these lines was treated separately by normalising their adjacent continuum to unity. Finally, for the most interesting epochs the high energy part of the spectrum, and in particular the Fe~\textsc{xxvi} Ly$\alpha$ transition (1.778 \AA; 6.973 keV), was also studied.  For clarity reasons, a slight Gaussian smooth of one pixel was used when representing the X-ray data. The data analysis was performed using \textsc{molly, xspec} and custom routines developed under \textsc{python}.

\begin{figure*}
    \centering
    \includegraphics[width=9truecm,height=16truecm]{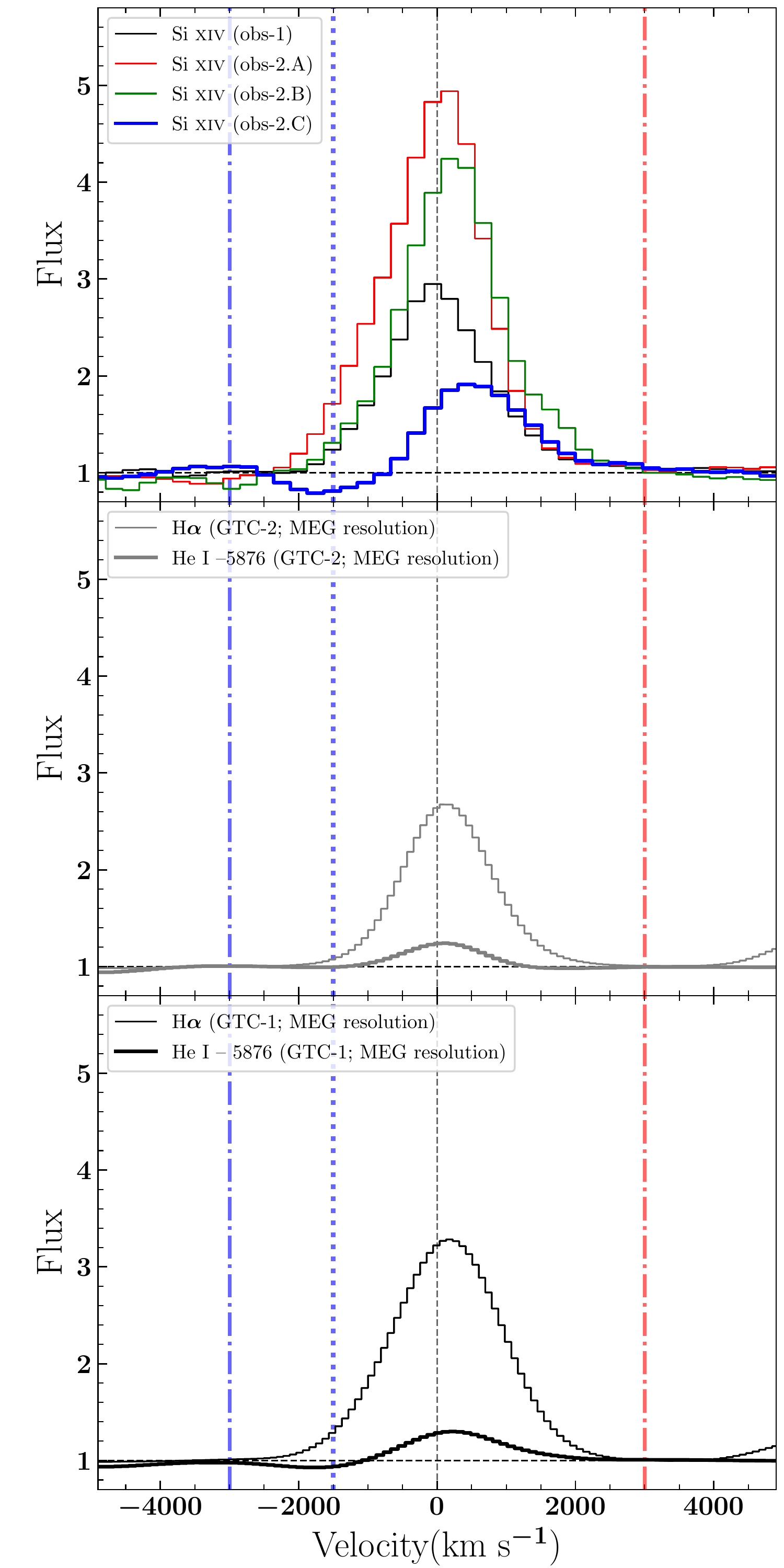}
    \includegraphics[width=9truecm,height=16truecm]{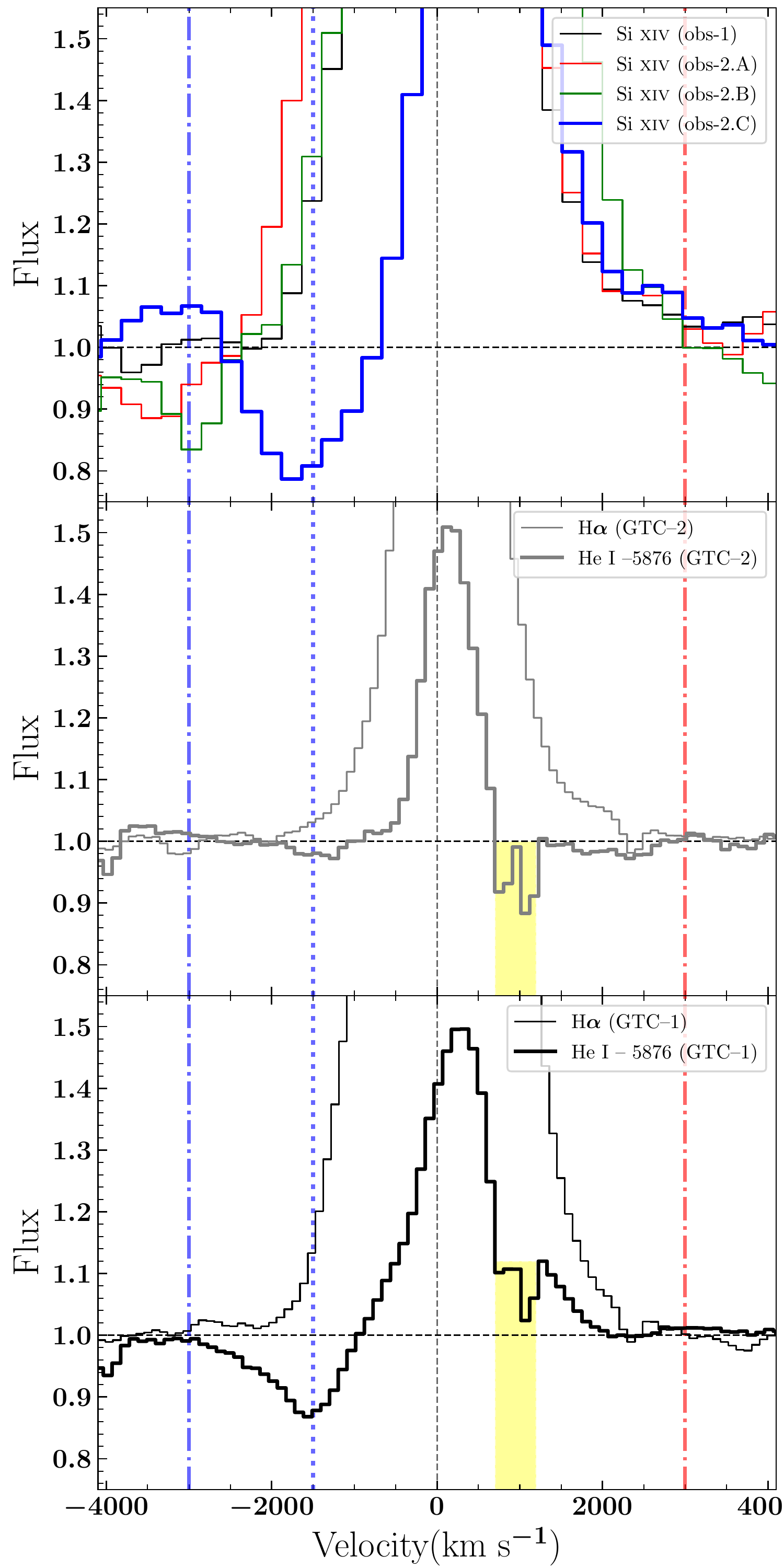}
    \caption{\textit{Chandra} spectra of Si~\textsc{xiv} (top panels) compared with GTC spectroscopy in the  He~\textsc{i} and \ha\ regions (mid and bottom panels). The right panels are zoomed version of the left ones aimed at highlighting the blue-shifted absorption features. Dotted vertical lines mark the core velocity (–1500 \kms) of the He~\textsc{i} absorption component, while the dotted-dashed vertical lines mark (approximately) its blue-edge velocity (blue; –3000 \kms) and the corresponding receding velocity (red; 3000 \kms ). }
    \label{fig:low-energy}%
\end{figure*}    
 
\subsection{\textit{Chandra} observation-1}
In agreement with  \citetalias{King2015}, no strong wind features are detected during obs-1 (although spec-1.4 at high flux shows features compatible with weak wind signatures). Given the lack of conspicuous wind-related absorptions and simultaneous \textit{INTEGRAL} and GTC observations, we decided to use the average spectrum for the entire obs-1 to be compared with the different sets of data derived from obs-2, which will be the main focus of our analysis (see below).  We note that a very significant count-rate drop was observed in the last spectrum (i.e. spec-1.10), which resulted in a significantly lower S/N. This is consistent with the flux drop observed in the optical light curves presented in \citet[][see fig. 1]{Kimura2016} around that time. Therefore,  even though \mbox{GTC-1} started merely 28 minutes after obs-1 ended, this strong flux variability suggest that the properties of the source might be very different between obs-1 and GTC-1, and thus we will treat these windows as completely independent datasets (see Sec. \ref{discussion}).
 
Interestingly, strong optical wind signatures are present during GTC-1.  In particular, \hei\ shows a P-Cyg profile with a blue absorption component at $\sim 1500$ \kms (dotted vertical line) and a blue-edge velocity of $\sim 3000$ \kms (blue, dotted-dashed vertical line). The \ha\ line profile is redshifted but shows a prominent blue wing in emission (Fig. \ref{fig:low-energy}, right panel). A detailed view of the \ha\ evolution throughout  this observation can be found in Fig. 15 of \citetalias{MataSanchez2018}.  

 \subsection{\textit{Chandra} observation-2}
Obs-2 was fully covered by \textit{INTEGRAL} data (Fig. \ref{fig:data}), where large flux variations become evident. In particular, the last two spectra (spec-2.8 and -2.9) include two large flares, with the hard X-ray flux increasing by more than three orders of magnitude with respect to the beginning of the observation. P-Cygni profiles are witnessed during these two epochs, the strongest being those during spec-2.8, in correspondence with the most luminous flares. In order to increase the S/N of the spectra, we divided obs-2 in three intervals based on the \textit{INTEGRAL} light curve, which is much less affected by intrinsic absorption than soft X-rays and therefore provides a more realistic estimation of the true luminosity of the system (i.e. unabsorbed; see Section \ref{intrinsic}). Thus, obs-2.A includes spec-2.1 to -2.4 (low-flux), obs-2.B  covers spec-2.4 to -2.7; (plateau) and obs-2.C includes spec-2.8 and -2.9 (flares). 
Fig. \ref{fig:low-energy} (left panel) shows how the equivalent width of the Si~\textsc{xiv} emission line decreases across obs-2, following the increase in hard X-ray luminosity. In addition, obs-2.C  is characterised by strong P-Cyg features with blue-shifted absorptions reaching values $\sim 20$ per cent below the continuum level. Both the core velocity of the absorption component and its blue-edge velocity are comparable to those observed in \hei\ in GTC-1 (Fig. \ref{fig:low-energy}; right panel). The blue emission line wings observed in obs-2.A and obs-2.B (and also in obs-1) reach similar velocities than the blue-edge velocities determined in GTC-1 (i.e. \hei) and obs-2.C (dotted-dashed, blue lines). This is also true for the red-edge of the emission line profile, which roughly match the velocity observed in the blue-edge of the P-Cyg absorptions (dotted-dashed lines; red when taken positive).  

\begin{figure}
    \centering
    \includegraphics[width=9truecm]{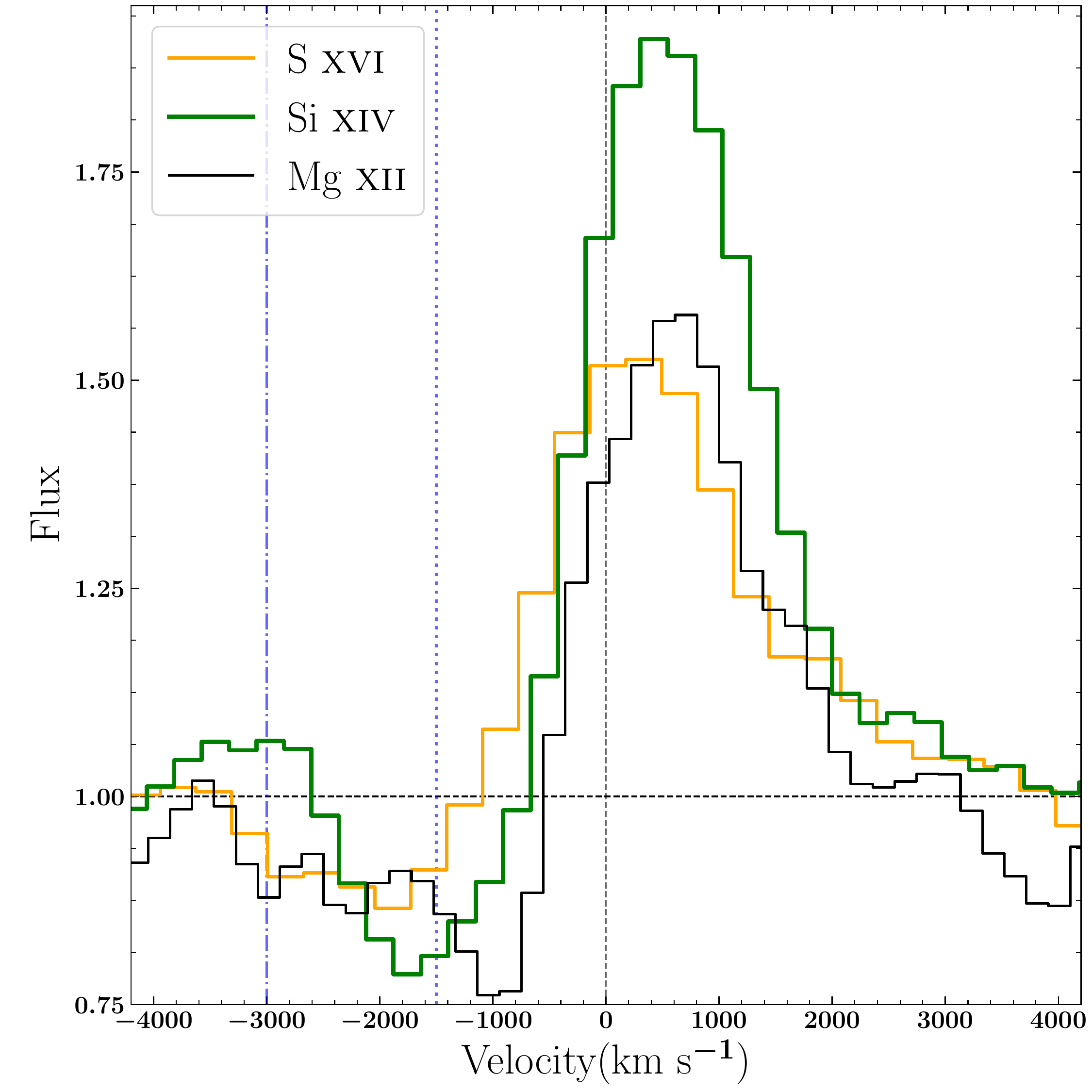}
    \caption{Detection of P-Cyg profiles in the S \textsc{xvi}, Si \textsc{xiv} and Mg \textsc{xii} emission lines during obs-2.C. The purple dotted-dashed and dotted vertical lines mark the same velocities as in Fig. \ref{fig:low-energy}.}
    \label{fig:P-Cygs}%
\end{figure}

The P-Cyg profile observed in Si~\textsc{xiv} is also detected in S~\textsc{xvi} and Mg~\textsc{xii} with roughly similar properties, although slight variations in the core and blue-edge velocities of the blue-shifted absorption components might be present (Fig. \ref{fig:P-Cygs}; see Sec. \ref{discussion}).

\begin{figure*}
    \centering
    \includegraphics[width=9cm]{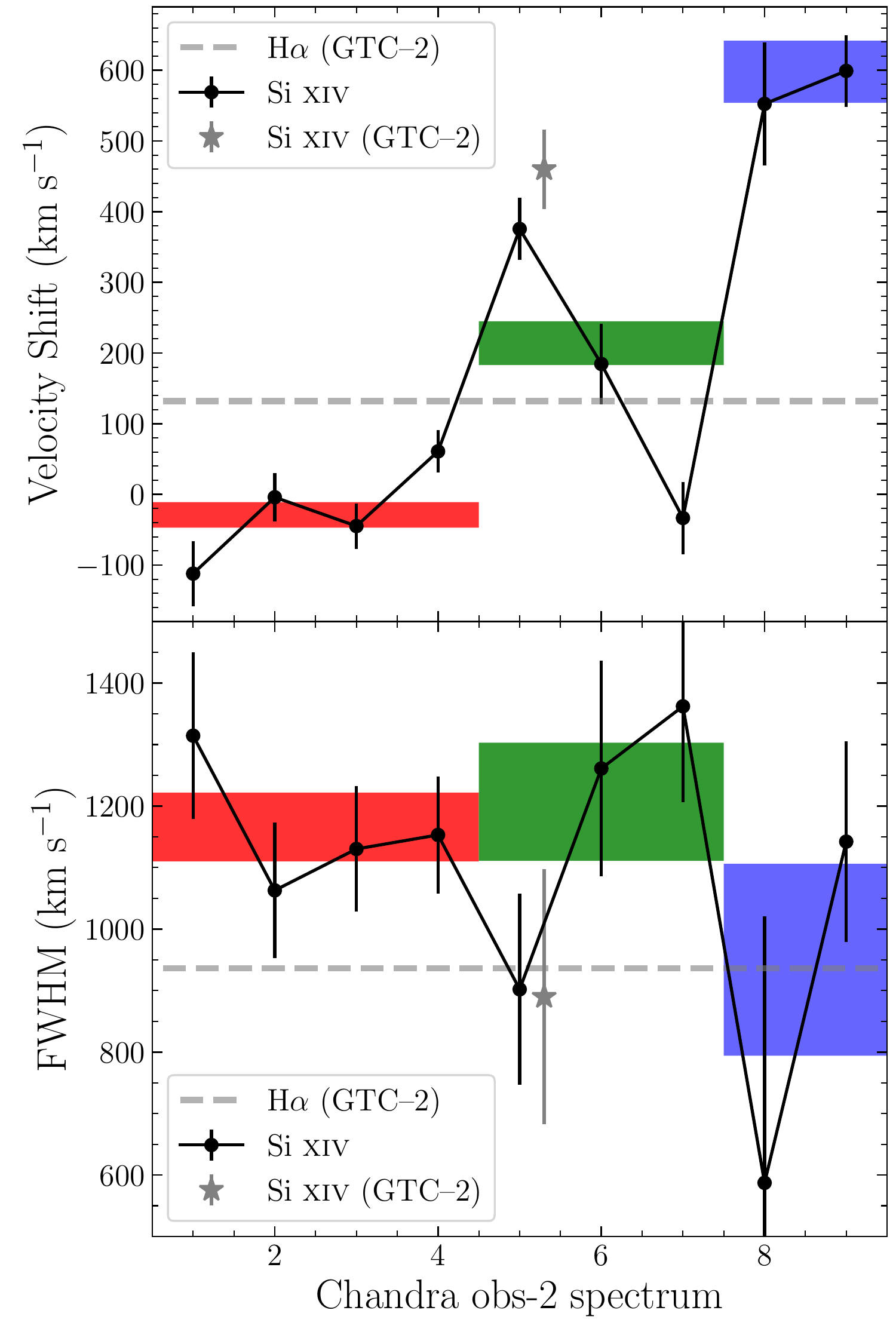}
    \includegraphics[width=9cm]{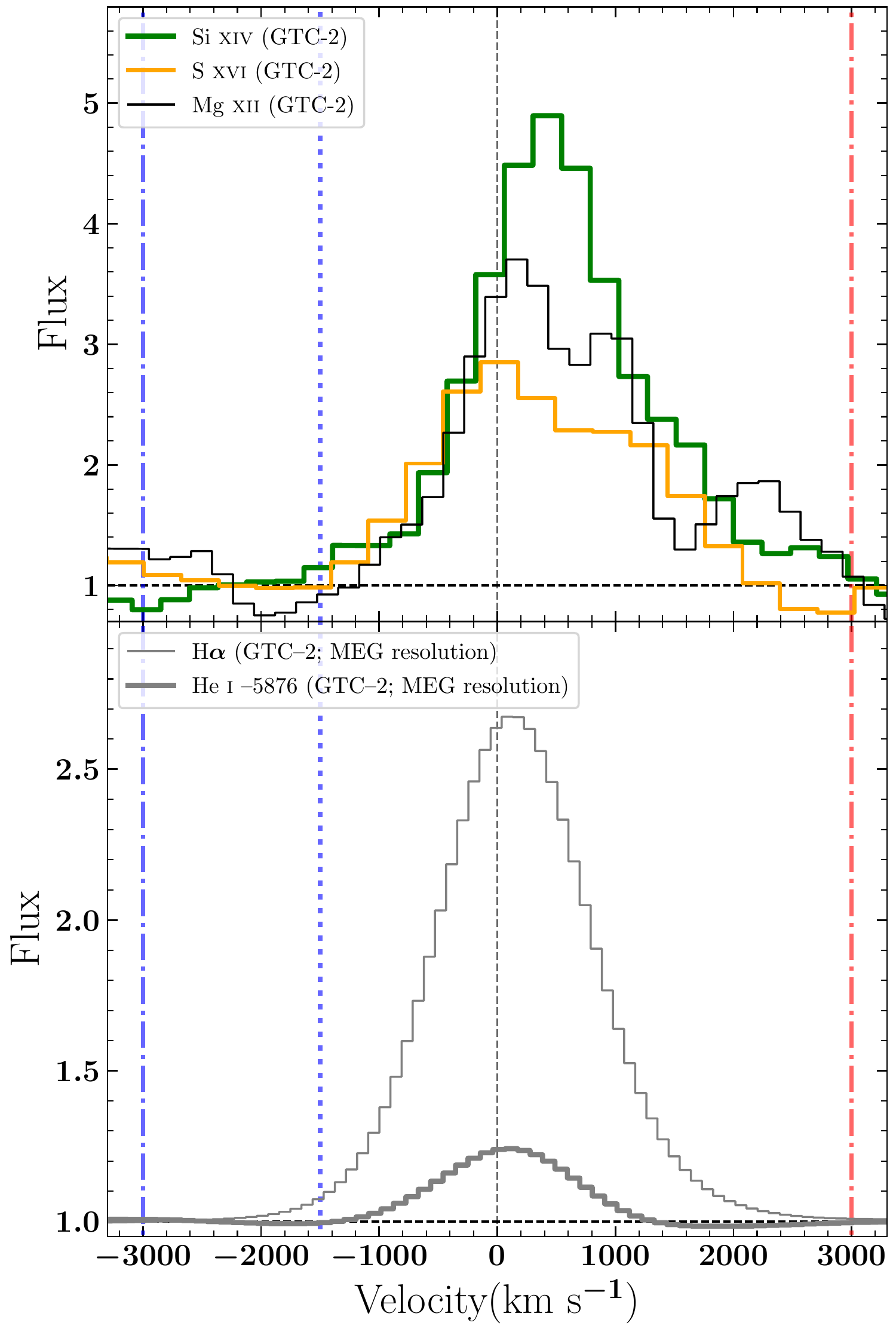}
\caption{Emission line properties during obs-2.  Left panels: velocity offset of Si \textsc{xiv} (top-left). The coloured horizontal bands mark the results for obs-2.A, obs-2.B and obs-2.C (with their velocity width representing $\pm 1\sigma$  errors). The colour code is the same as in Fig \ref{fig:data}. The evolution of the FWHM is presented in the bottom-left panel, where the horizontal dashed lines indicate the FWHM of \ha\ during GTC-2. The Si \textsc{xiv} measurement obtained during GTC-2 (i.e. top-right) is indicated by a star. Right panels: Emission line profiles (in velocity scale) of Mg \textsc{xii},  Si \textsc{xiv}, S \textsc{xvi} (top-right); \ha\ and He~\textsc{i} (bottom-right) during GTC-2. For an easier visual comparison, the velocity resolution of the optical lines has been degraded to that of \textit{Chandra}. The vertical lines are the same as in Fig. \ref{fig:low-energy}.}
       \label{fig:gfit}%
\end{figure*}    

\subsubsection{Emission line variability} 
\label{gfit}
Si \textsc{xiv} is the strongest emission line in the soft part of the spectrum and shows strong variability throughout obs-2, including blue-shifted absorptions during obs-2.C. In order to characterise this variability we performed a Gaussian fit to the line profile of the $9\times$3.2 ks spectra, with the velocity offset (from the rest wavelength), the width (corrected from the MEG velocity resolution), and the height of the Gaussian as free parameters. The results are presented in Fig. \ref{fig:gfit} and Table \ref{tab:fits}. The centroid of Si \textsc{xiv} is observed to shift towards larger velocity values as the hard X-ray activity increases. In particular, the largest velocity offsets are measured in the spectra that include flaring activity (see Fig. \ref{fig:data}): spec-2.5, -2.6 and particularly spec-2.8 and -2.9. Contrastingly, the lowest values are found at low X-ray fluxes (i.e. obs-2.A) and also in spec-2.7 during the plateau phase (but without clear flaring events present). Also,  S~\textsc{xvi} and  Mg~\textsc{xii}   show significant variability throughout obs-2. However, their worse S/N results in larger error bars when fitting the $9\times$3.2 ks spectra. Nonetheless, when using the three spectra of the obs-2 intervals we obtained velocity offsets that are consistent with the trend seen in Si \textsc{xiv} (see Table \ref{tab:fits}). We note that although the theoretical (absolute) wavelength accuracy at Si~\textsc{xiv} (6.184 \AA) varies between 290 and 530 \kms (HEG and MEG), it has been empirically found to be as good as $\sim$ 30 \kms for both HEG and MEG data \citep{Ishibashi2006}. Nevertheless, the relative wavelength accuracy between observations is expected to be in the range of $\sim$ 50--100 \kms.
The above indicates that the observed trend is real and that significant redshifts are associated with the epochs dominated by luminous hard X-ray flares.

The FWHM of Si \textsc{xiv} is observed to be  $1166 \pm57$, $1207\pm 96$  and $950\pm156$ \kms for obs-2A, -2B, and -2C, respectively. S~\textsc{xvi} and  Mg~\textsc{xii} show roughly consistent values (see Table \ref{tab:fits}). The narrower lines observed during obs-2C are in agreement with the detection of P-Cyg profiles. 
Table \ref{tab:fits} also reports the fits for the three Ly$\alpha$ transitions for the entire obs-1 and obs-2. In agreement with \citetalias{King2015}, we obtain small velocity offsets in the range of $\sim \pm$ 100 \kms for both observations, although values tend to be redshifted for obs-2, as expected given the presence of P-Cyg profiles. The typical FWHM for both observations is $\sim 1200$ \kms.

\begin{table*}[ht!]
        \centering
        \caption{Gaussian fits to the Ly$\alpha$-like emission lines}
        
        \begin{threeparttable}
        \begin{tabular}{l c c c c c c c}
        
                \hline
                \textit{Chandra} Window & Mg \textsc{xii} offset & Mg \textsc{xii} FWHM & Si \textsc{xiv} offset & Si \textsc{xiv} FWHM & S \textsc{xvi} offset & S \textsc{xvi} FWHM \\
                  & (\kms) & (\kms) & (\kms) & (\kms) & (\kms) & (\kms) \\
                \hline
                \hline
                Obs-1  & $-102 \pm 28$ &$1302 \pm76$ & $-1 \pm 24$ &  $1241 \pm77$ & $-4 \pm 67$ & $1021 \pm268$  \\
                Obs-2  & $136 \pm 24$ &$1041 \pm70$ & $153 \pm 16$ &  $1168 \pm52$ & $25 \pm 44$ & $1005 \pm188$  \\
                \hline
                Obs-2.A  & $9 \pm 21$ &$869 \pm68$ & $-28 \pm 18$ &  $1166 \pm57$ & $-85 \pm 52$ & $1049 \pm213$  \\
                Obs-2.B & $172 \pm 48$ &$1199\pm 130$& $213\pm 31$ & $1207\pm 96$ & $64\pm 86$ & $1152\pm 342$  \\
                Obs-2.C & $664 \pm 88$ &$1236\pm247$ & $598 \pm 44 $& $950\pm156$ & $322 \pm 110$ & $680\pm634$  \\               
                \hline
                GTC-2 & $384\pm106$ &$1294 \pm 273$& $460\pm56$ & $889\pm207$ & $312\pm220$ & $1349\pm776$  \\
                
                \hline
        \end{tabular}
        \label{tab:fits}
        
        \end{threeparttable}
\end{table*}

\subsubsection{Strictly simultaneous data} 
Observation GTC-2 was partially covered by \textit{Chandra} spec-2.5 and spec-2.6. In order to perform a more direct comparison we computed the \textit{Chandra} spectrum during the 2.3 ks simultaneous with GTC-2.  The associated hard X-ray flux during GTC-2  is one order of magnitude lower than the peak of the flares in obs-2.C. However, it includes a smaller flare that occurred during the transition between spec-2.5 and spec-2.6 (Fig. \ref{fig:data}). Fig. \ref{fig:gfit} (right panels) shows the line profiles of the X-ray and optical lines included in our analysis.  None of the emission lines show P-Cyg profiles, but they are all significantly redshifted (see Table \ref{tab:fits}). In particular, Si \textsc{xiv} is roughly symmetric and centred at $\sim 400$ \kms, while the optical lines are centred at $\sim 140~\kms$ (\citetalias{Munoz-Darias2016}). S~\textsc{xvi} and  Mg~\textsc{xii} show nominal redshifts consistent with those of Si \textsc{xiv}. However, these lines are more complex and non-symmetric, with a secondary red component clearly present and a main emission line peak that is only slightly redshifted (similar to that of the optical lines).    

The FWHM of \ha\ during GTC-2 is $\sim 940$ \kms (varying in the range of 875 --993 \kms; \citetalias{MataSanchez2018}), which is lower than observed in obs-2.A and -2.B (1100--1300 \kms). However, when computing the FWHM of Si \textsc{xiv} during GTC-2, we obtain $889 \pm 207$, which is consistent with that measured in the optical emission lines (see also Fig \ref{fig:gfit}).

\subsubsection{High-energy spectrum}
\begin{figure*}
    \centering
    \includegraphics[width=18truecm]{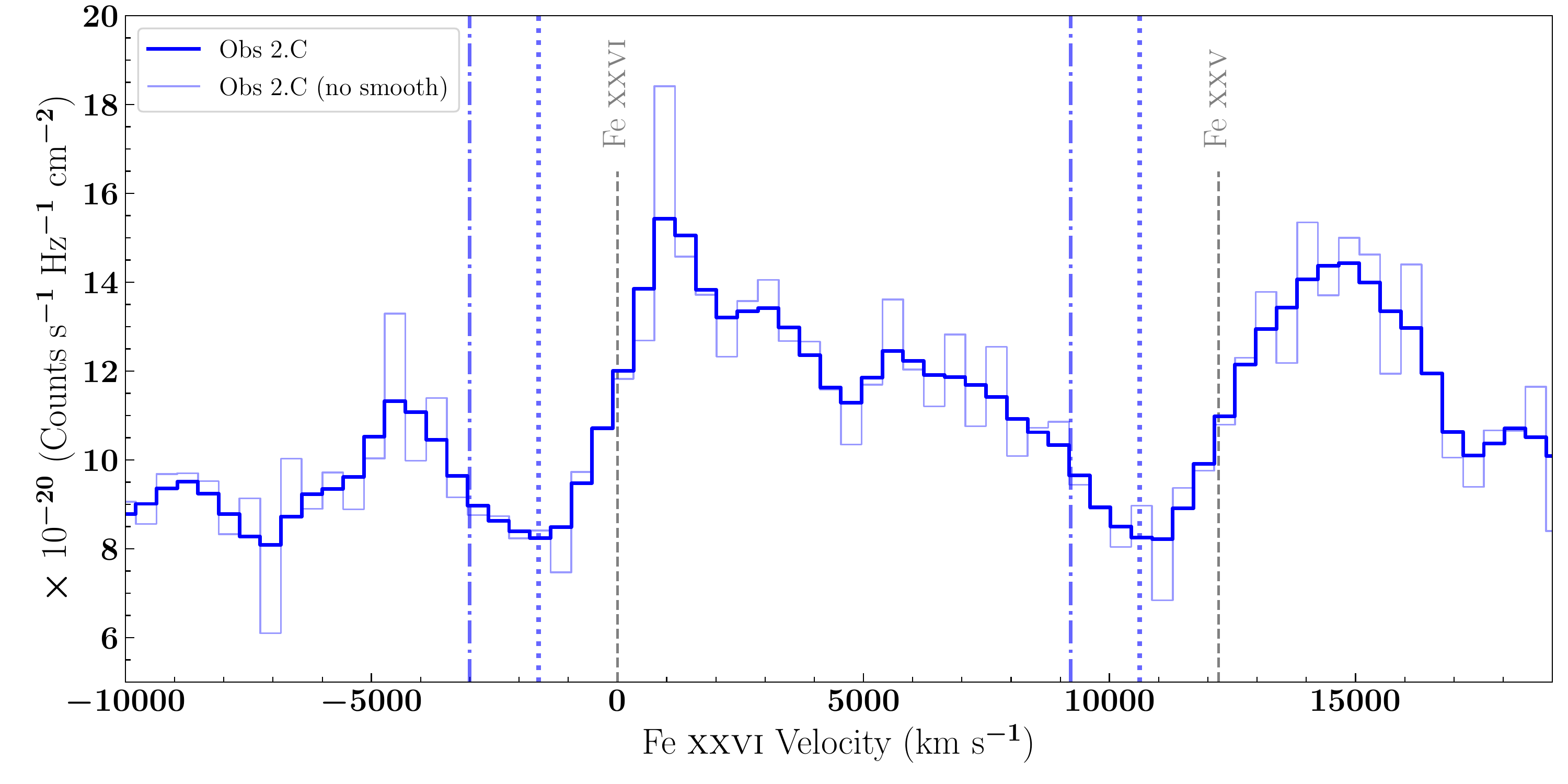}
    \caption{Blue-shifted absorptions in Fe~\textsc{xxvi} and Fe~\textsc{xxv} during obs-2.C. The purple dotted-dashed and dotted vertical lines mark the same velocities as in Fig. \ref{fig:low-energy}.}
    \label{fig:HEG}%
\end{figure*}

The \textit{Chandra} high-energy spectrum is significantly more complex than the soft one. A detailed analysis of this spectral region is beyond the scope of this work. However, we note that blue-shifted absorptions are present during obs-2.C (see also \citetalias{King2015}). Figure \ref{fig:HEG} shows the Fe~\textsc{xxvi} (Ly$\alpha$) transition together with Fe~\textsc{xxv}.  These components are broader than those observed at lower energies but their core velocities match (rather precisely) that of \hei\ (and, consequently, the soft X-ray transitions during obs-2.C; Fig. \ref{fig:low-energy}). The continuum of the high-energy spectrum is more complex than that of the soft X-rays lines, which hampers a clear identification of the blue-edge velocity. However, it might be still consistent with the $\sim 3000$ \kms observed in the optical and soft X-ray emission lines (as we discuss below).

\section{Discussion}
\label{discussion}
The black hole transient V404 Cygni displayed strong X-ray (\citetalias{King2015}) and optical (\citetalias{Munoz-Darias2016}) accretion disc wind signatures during the most active part ($\sim 12$ days) of its 2015 outburst.  In this work, we have analyse the only contemporaneous  observations  (including simultaneous data) taken with \textit{Chandra} and GTC over two consecutive days within this activity period. Although our study includes recombination lines of species with largely different excitation potentials, the kinetic properties of the wind are remarkably similar. In particular, we find a common characteristic velocity (as derived from the core of the optical and X-ray blue-shifted absorption lines) of $\sim 1500$ \kms and a blue-edge velocity (a proxy for the wind terminal velocity) $\gtrsim 3000$ \kms (e.g. Fig. \ref{fig:P-Cygs}). However, while we do observe strictly simultaneous redshifts in both the optical and the X-ray emission lines (Fig. \ref{fig:gfit}), the P-Cyg features (i.e. blue-shifted absorptions) are observed at different times and hard X-ray fluxes.  

\subsection{Observational properties of the wind}
As discussed in \citetalias{Munoz-Darias2016} and \citetalias{MataSanchez2018}, conspicuous optical P-Cyg profiles are only observed at low optical (and hard X-ray) fluxes.  Observations characterized  by a low average flux but including flaring activity are particularly revealing, since they show how the optical P-Cyg profiles briefly disappear and reappear in correspondence to flares (e.g. Fig 2 in \citetalias{Munoz-Darias2016}). This behaviour is accompanied by a sudden increase in the intensity ratio of the higher excitation emission lines (e.g. He \textsc{ii} and N \textsc{iii}) to the low excitation ones (e.g. He \textsc{i}, \ha) during the flares. This indicates that (at least during low-flux epochs) the wind is permanently active, but its detection via P-Cyg profiles is most likely driven by ionisation effects. As a matter of fact, throughout the entire outburst the optical P-Cyg profiles are solely detected when these line ratios indicate a low-ionisation optical spectrum, while other observables (e.g. redshifted lines, broad emission line wings) strongly suggest that the outflow was likely present during most, if not the entire event (\citetalias{Munoz-Darias2016}).  

Observation GTC-1 shows strong optical P-Cyg profiles as well as an optical flare, when the wind signatures briefly disappear (see Fig. 15 in \citetalias{MataSanchez2018}). This observation occurs during an optical flux drop, that  starts right at the very end of \textit{Chandra} obs-1 (day $\sim 7.5$ in Fig. 1 in \citealt{Kimura2016}). Contrastingly, both the optical flux and, in particular, the ionisation state of the optical emitting accretion disc (traced by line ratios) are higher during GTC-2 and P-Cyg profiles are not observed. 

In X-rays, a rather opposite behaviour is witnessed: P-Cyg profiles are only present during luminous flares, and these are seen in emission lines indicating mid (e.g. Si \textsc{xiv}) to high (e.g. Fe \textsc{xxvi}) ionisation parameters. \citetalias{King2015} suggested that the terminal velocity of the X-ray wind, as measured from the blue-edge velocity of the P-Cyg absorption, increases with the ionisation potential of the atomic species from $\sim1500$ to $4000$ \kms. However, we note that this is a particularly difficult measurement that requires a good determination of the continuum. Our results shows that while the high energy lines  (Fig. \ref{fig:HEG}) have broader absorptions profiles than the lower excitation ones, core velocities of $\sim 1500$ \kms and blue-edge velocities of $\sim 3000$ \kms are roughly compatible with all the optical and X-ray P-Cyg profiles observed (dotted and dashed-dotted lines in Figs \ref{fig:low-energy}, \ref{fig:P-Cygs}, \ref{fig:HEG}). Nonetheless, terminal velocities of $\sim 4000$ \kms are also consistent with the blue-edges of Fe \textsc{xxvi} and Fe \textsc{xxv} during obs-2.C, as proposed by \citetalias{King2015}.        

\subsection{Wind variability and actual luminosity of V404~Cygni }
\label{intrinsic}
We have discussed how the observational properties of the optical and X-ray wind change with time, following the large variations in the observed luminosity. However, detailed X-ray spectral analysis favours the presence of variable X-ray absorption, which might reach very high values (\nh $\sim 10^{23-24}$ cm$^{-2}$; e.g. \citealt{Motta2017a, Motta2017b}), significantly exceeding the Galactic extinction along the line of sight (\nh $\sim 10^{22}$ cm$^{-2}$). Variable X-ray absorption was also reported for the 1989 outburst (e.g. \citealt{Zycki1999}) and can be roughly accounted by the estimates of the wind mass outflow rate derived from the optical spectra (\citetalias{Munoz-Darias2016}; \citealt{Casares2019}). It is worth noting that this behaviour, namely, conspicuous outflows and variable absorption, is not exclusive of V404 Cygni  and has, in fact, been seen in other systems (a relatively long orbital period seems to be a common feature; see e.g. \citealt{Munoz-Darias2018, Koljonen2020}). While this large and variable \nh\ is expected to have a strong impact on the soft X-ray variability, it is less clear how much it affects the hard X-ray light curve. For instance, assuming a simple power-law spectrum, the \textit{INTEGRAL} count-rate (25--200 keV) is expected to drop by $\sim 5$--$10$ percent (intrinsic power-law index in the range of 1.5--3) when \nh\  increases from $10^{22}$ to  $5\times 10^{24}$ cm$^{-2}$.  The same \nh\ variation would produce a drop of four order of magnitudes in the 2--10 keV band. Incidentally, this level of variability is present in the soft X-ray light curve of the outburst (e.g. that from \textit{Swift}, see e.g. \citetalias{Munoz-Darias2016} and \citealt{Motta2017a}). Still, a fairly similar variability level  ($\sim$ three orders of magnitude) is also present in the \textit{INTEGRAL} light curve (see also Fig. \ref{fig:data}), as well as in the radio emission (\citetalias{Munoz-Darias2016}; see also \citealt{Tetarenko2017, Tetarenko2019}).  This indicates that an important fraction of the hard X-ray variability is most likely intrinsic and, thus, it is expected that the soft X-ray light curves have also a significant intrinsic variability component.  A similar conclusion was drawn by \citet{Hynes2019} from the study of the X-ray/optical correlations (see also \citealt{AlfonsoGarzon2018}) and the behaviour of the hard X-ray light curve, which often displays a harder-when-brighter trend during flares. These hard flares might be explained by a strong reflection component from a jet-illuminated disc, a scenario that would account for the hot seed photons derived in some spectral fits (\citealt{Walton2017}; see also \citealt{Roques2015, Natalucci2015}). However, very large \nh\ values might also account for this problem (see e.g. \citealt{SanchezFernandez2017}). 

On a related note, using soft X-ray data of GRS~1915+105 during a phase that is also characterised by high levels of intrinsic absorption, \citet{Neilsen2020} found that the flaring behaviour can be explained by a combination of enhanced intrinsic activity and reduced absorption by a similar factor (3--4). The latter is likely (to some extent) a direct consequence of ionisation effects on the obscuring material, which at least for the case of V404 Cygni seems reasonable to assume that must be linked somehow to the outflow.

Finally, we note that a large mass outflow rate, such as that inferred for V404 Cygni, might also drive violent intrinsic variability as a result of instabilities in the accretion rate (e.g. \citealt{Neilsen2011}, \citetalias{Munoz-Darias2016}).    


\subsection{A multi-phase accretion disc wind}
\label{wind_structure}
As discussed above, the kinetic properties of the X-ray and optical outflows are remarkably consistent with each other. However, P-Cyg profiles -- the most convincing evidence for the presence of winds -- are not observed simultaneously, nor at the same range of luminosities. There are several scenarios that may explain this behaviour (see illustrative sketch\footnote{created by Gabriel Pérez, multimedia team, Instituto de Astrofísica de Canarias.} in Fig. \ref{fig:sketch}):

\begin{figure*}
    \centering
    \includegraphics[width=18truecm]{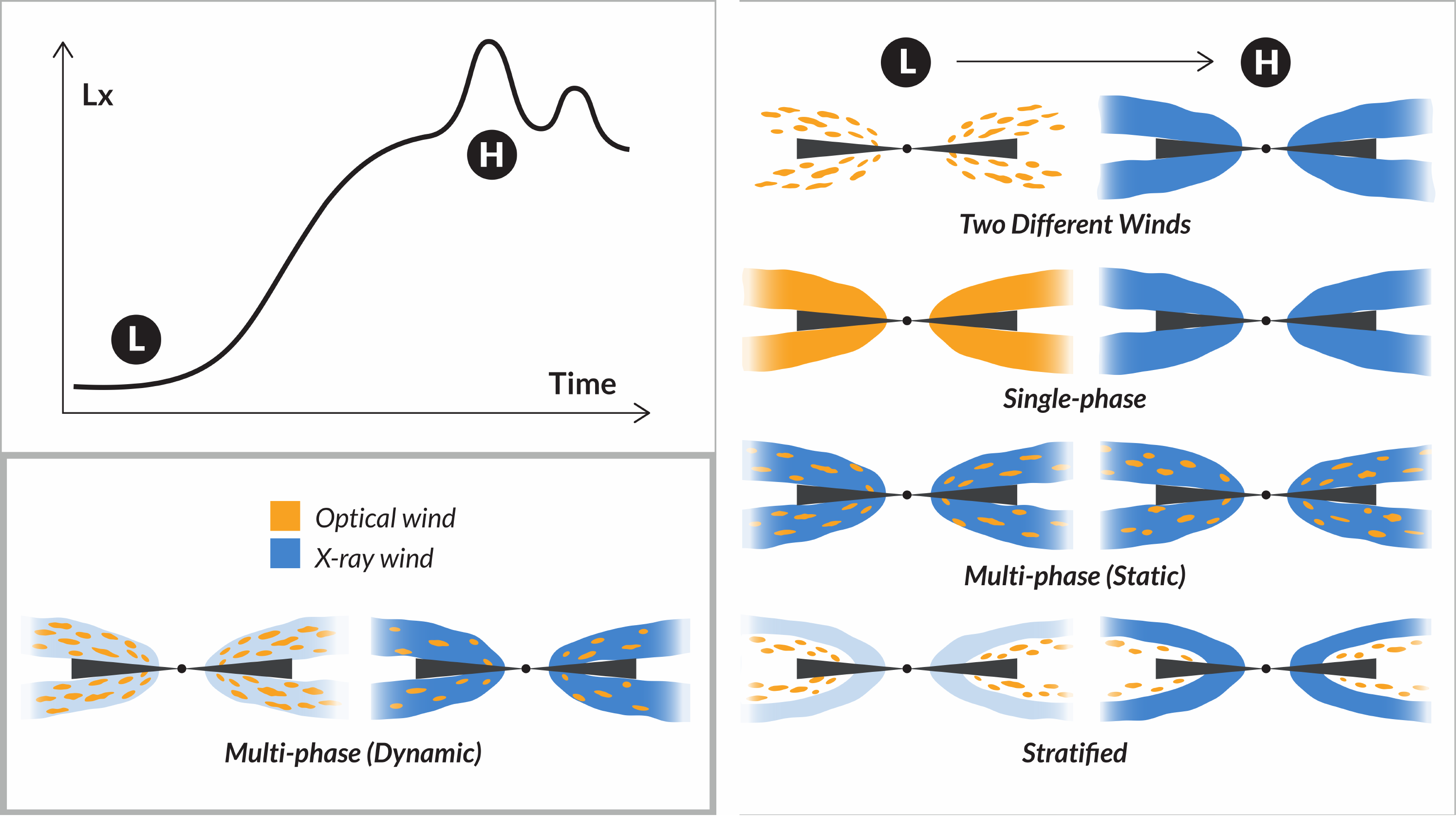}
    \caption{Illustrative sketch depicting the different possibilities discussed in Sec. \ref{wind_structure} on the relation between the optical and X-ray winds as the hard X-ray luminosity varies from low (L) to high (H) values. We find that a multi-phase, dynamic scenario is favoured by the observations.}
    \label{fig:sketch}%
\end{figure*}

\paragraph{(i) Two different winds.}  There are two independent outflows that are triggered at different luminosities. The similarity between their kinetic properties would be merely casual or somehow related to the physical properties of the system (e.g. physical size), but not to the nature of the wind. It might be also expected that winds are launched from different regions within the disc. While it is difficult to fully rule out this scenario, the remarkably similar emission line properties derived from the simultaneous observations (assuming both X-ray and optical lines are formed in the wind; see below) implies significant fine-tuning to keep both outflows as totally independent phenomena.
\paragraph{(ii) A single-phase wind.} The wind continuously responds to the changes in the properties of the irradiating luminosity (e.g. amount and spectral shape), which is expected to trigger changes in the ionisation parameter of the outflowing gas, and thus in its observability across the electromagnetic spectrum. Hence, there is either an X-ray wind or an optical wind, but not coexisting phases (e.g. hot and cold) of the outflow. This scenario would be ruled out if the cold wind is always active, as optical observations seem to indicate. For instance, excluding epochs with P-Cyg profiles, V404 Cygni showed optical emission lines systematically broader during the 2015 outburst than during quiescence. This is at odds with what is typically observed in other sources, that display  narrower (and mostly double-peaked) lines during outburst than in quiescence, as they arise in outer disc regions characterised by lower velocities. In addition, optical lines are systematically red-shifted during the first $\sim 10$ days of the outburst (see \citetalias{MataSanchez2018, Munoz-Darias2016}). As a matter of fact, during the simultaneous \textit{Chandra}/GTC window (GTC-2), the optical and X-ray lines are both redshifted and their breadths are consistent with each other (see Fig. \ref{fig:gfit}), suggesting that they both arise in (the same) wind, which would rule out this scenario.
\paragraph{(iii)  A multi-phase static wind.} The wind has both hot (X-ray) and cold (optical) phases, but their physical properties are the same across the outburst. This is ruled out as we see variability in the observational properties of the wind as a function of the luminosity.
\paragraph{(iv)  A stratified wind.}
The X-ray and optical winds coexist in time, but their balance and observability vary according to physical changes in the system and/or the irradiating spectrum.  The ionisation parameters of the X-ray and optical emitting regions of the outflow are expected to be significantly different, namely, $\xi_\mathrm{X} >> \xi_\mathrm{V}$. From this, we can simply derive:

\begin{equation}
\frac{L}{n_\mathrm{X}r^{2}_\mathrm{X}} >>\frac{L}{n_\mathrm{V}r^{2}_\mathrm{V}} 
,\end{equation}

where we assume that both the hot and cold ejecta see the same luminosity ($L$), while $n_\mathrm{X}$ and $n_\mathrm{V}$ are the densities for the X-ray and optical emitting gas, and $r_\mathrm{X}$ and $r_\mathrm{V}$ the corresponding distances from the irradiating source.  Thus, if the density is constant throughout the wind (i.e. $n_\mathrm{X} \approx n_\mathrm{V}$), we would end up with a stratified wind ($r_\mathrm{V}$ > $r_\mathrm{X}$; i.e. the optical wind is further away from the black hole).  From the simulations of optically thin nebulae by \citet{Kallman1982}, it seems reasonable to assume that, in order to detect the spices present in the optical and X-ray spectra, there is at least a three orders of magnitude difference between their corresponding ionisation parameters, that is, $\xi_\mathrm{X} \gtrsim 10^3 \xi_\mathrm{V}$. This implies $r_\mathrm{V} \gtrsim  30 r_\mathrm{X}$.  While this case (constant density) remains useful for illustrative purposes, it is probably unrealistic since the wind is expected to diffuse into the interstellar medium with a given opening angle. This implies that density naturally decreases with the distance ($r$) to the source. For instance, if we repeat the above exercise for a conical wind that diffuses uniformly (i.e. no clumps) following a $r^{-1}$ density law, we obtain an even more restrictive solution of $r_\mathrm{V} \gtrsim  10^{3} r_\mathrm{X}$. It is important to bear in mind that the optical wind is observed to respond to the variations of the X-ray source on time scales $\lesssim 80$ seconds, which places the ejecta within the accretion disc outer radius ($\sim$ 30 light-seconds; i.e. $9\times10^{6}$ km) or very close to it (\citetalias{Munoz-Darias2016}). Assuming that  the wind velocity ($\sim 1500-3000$ \kms) traces the escape velocity at the launching region, a radius of $\sim 1$ light-second is obtained, which would set a lower limit for the distance to the X-ray wind, implying that the optically emitting ejecta would need to be placed at $\sim$ 30--1000 light-seconds.  On the other hand, \citetalias{King2015} derived a radius of a few times $10^{6}$ km for the formation of some of the X-ray lines -- which, if arising in the wind, would put the optical outflow even further away.  All the above strongly suggest that stratification, all by itself, cannot explain the observables. 


\paragraph{(v)  A multi-phase dynamic wind.}  

Finally, one can also consider that the X-ray and optical winds coexist not only in time but also in space (i.e. $r_\mathrm{V} \approx r_\mathrm{X}$). 
Thus, $\xi_\mathrm{X} \gtrsim 10^3 \xi_\mathrm{V}$ implies $n_\mathrm{V} \approx 10^3 n_\mathrm{X}$. In this case, the wind would be a combination of optically emitting denser and cooler blobs that are embedded in thinner and hotter gas regions which,  under isobaric conditions follow $n_\mathrm{V}T_\mathrm{V} = n_\mathrm{X}T_\mathrm{X}$, with $T_\mathrm{V}$ and $T_\mathrm{X}$ the temperatures of the optical and X-ray emitting gas, respectively. This configuration is often found in nebulae in the interstellar medium (e.g. \citealt{Mckee1977}), albeit at lower densities. This is also commonly applied to active galactic nuclei in order to explain their co-spacial outflows, which, similarly to the case of V404 Cygni, cover several orders of magnitude in ionisation parameter (e.g. \citealt{Krolik1981}).  As a matter of fact, theoretically oriented photoionisation studies (e.g. \citealt{Chakravorty2013, Bianchi2017, Gatuzz2019}) have shown that, for typical X-ray binary spectral energy distributions, multiple gas phases can be  stable in pressure equilibrium. Even if this condition is not satisfied (e.g. as a result of significant radiation pressure from the X-ray source) alternative solutions allow for largely different ionisation parameters within a single slab of gas (\citealt{Stern2014b,Stern2014}).  
Thus, if the irradiating X-ray source is strongly variable, as it is the case in V404 Cygni, this naturally becomes a very dynamic system, and therefore significant variability is expected in the observational properties of the outflow across the electromagnetic spectrum. This would be also true even if the variability is due to variable extinction (but see Sec. \ref{intrinsic}) as long as the main absorber is located between the X-ray source and the wind launching region (see e.g. \citealt{Motta2017a}). Multi-phase, clumpy, winds are also thought to be present in massive stars, where both observational and theoretical studies indicate that the wind perturbations originating the clumps occur from very inner regions of the wind (i.e. close to the launching region; see e.g. \citealt{Sundqvist2013} and references therein).  

We conclude that, although some degree of stratification might be present and it is probably not unexpected (e.g. predominately hotter gas to be present at the ionisation front in the very inner part of the wind), a multi-phase solution is able to explain the observables. Furthermore, we note that the similar kinematic properties found in both X-ray and optical winds in LMXBs (e.g. \citealt{Ponti2015, PanizoEspinar2022}), which show similar velocities in the range of a few hundred to a few thousands kilometres per second, together with the simultaneous detection of optical and near-infrared (\citealt{SanchezSierras2020}) as well as optical and ultraviolet (\citealt{CastroSegura2022}) winds in two systems, suggest that multi-phase accretion disc winds might be the norm in X-ray binaries.

\subsection{Some considerations on the wind launching mechanism}
In the above, we have discussed the properties of the wind once it has been launched.  However, the wind launching mechanism in X-ray binaries is an open question, and the case of V404 Cygni is no exception. The wind could have a magnetic origin (e.g. \citealt{Waters2018}), but the observed wind velocities are compatible with large launching radii (thousands of gravitational radii) and thus do not require this mechanism, which is able to produce outflows with large velocities if launched from the very inner accretion flow. Given that the source approached the Eddington luminosity during some phases of the outburst, a radiation driven wind (i.e. via Thompson scattering) might be unavoidable.  As a matter of fact, the large outflow mass (up to $\sim 100$ times the accreted mass) derived from the evolution of the nebular phase (\citetalias{Munoz-Darias2016}; see also \citealt{Rahoui2017}) that followed the most active phase of the event suggests a contribution from radiation pressure to the launch of the wind  \citep{Casares2019}. However, optical winds are observed at hard X-ray luminosities three orders of magnitude lower than the outburst peak. The most appealing mechanisms at these fainter epochs are Compton heated  (a.k.a thermal wind; \citealt{Begelman1983}) and line driven winds. Both mechanisms require of significant shielding against X-rays in order to not overionise the ejecta. As discussed above, a multi-phase, clumpy outflow could explain the lower ionisation parameter associated with the optical lines.  

Conspicuous P-Cyg profiles were seen in the optical lines across the first week of the outburst. During the first two days, the wind velocity (blue-edge) was significantly lower ($\sim 1500$ \kms) than that observed later in outburst ($\sim 3000$ \kms; e.g. GTC-1 in this paper; see Fig 1. in \citetalias{Munoz-Darias2016}). Interestingly, the observed luminosity of the system was lower in all the bands during these first two days of the event (\citealt{Kimura2016, Motta2017a}, \citetalias{Munoz-Darias2016}). Likewise, wind velocities were also lower ($\sim 1000$--$1500$ \kms) during the December 2015 secondary (and fainter) outburst (\citealt{Munoz-Darias2017}) and the (somewhat less extreme) 1989 outburst (\citealt{Casares1991}, \citetalias{MataSanchez2018}), suggesting a relation between the average luminosity of the system at a given stage of the outburst and the wind velocity. In this regard, it is an appealing approach to consider that the thermal wind, which is expected to have a slower response to luminosity changes, plays a role.  We note that the complex wind phenomenology seen in V404 Cygni suggests that more than one mechanism  might be contributing to the launch of the outflow. This probably should not come as a surprise since these mechanisms are unavoidable for certain physical conditions, and this would go in line with theoretical work that propose hybrid solutions to explain accretion disc winds in X-ray binaries (e.g. see \citealt{Higginbottom2020, Tetarenko2020} for recent studies).


\section{Conclusions}
We used contemporaneous (including simultaneous) X-ray and optical spectroscopic data, together with hard X-ray light curves, to study the main observational properties of the accretion disc wind of the black hole transient V404 Cygni. We have shown that the optical P-Cyg profiles observed at low hard X-ray fluxes are alike those seen in a number of X-ray transitions during luminous flares. Furthermore, strictly simultaneous data taken at intermediate hard X-ray fluxes show how the optical and the X-ray emission lines are both redshifted and have similar line profiles, indicating that they most likely arise in the same wind. All aspects considered, this study favours a scenario in which a dynamic, multi-phase accretion disc wind, whose observational properties dramatically change with the luminosity, was active during the entire 2015 outburst of V404 Cygni. New coordinated X-ray and optical/near-infrared observations of active LMXBs are highly encouraged to establish the predominantly multi-phase nature of X-ray binary winds, as the growing number of wind detections with similar properties (e.g. similar velocities) at largely different energies would suggest.

\begin{acknowledgements}
TMD acknowledges support from the Spanish Ministry of Science and Innovation via an \textit{Europa Excelencia} grant  (EUR2021-122010). TMD acknowledges support from the \textit{Consejería de Economía, Conocimiento y Empleo} of the Canary Islands and the European Regional Development Fund under grant ProID2020 010104. GP acknowledges funding from the European Research Council (ERC) under the European Union’s Horizon 2020 research and innovation programme (grant agreement No 865637). TMD acknowledges M.~Armas Padilla, J.~A.~Fern\'andez-Ontiveros and D. Mata-S\'anchez for their useful comments and suggestions. \textsc{Molly} software developed by Tom Marsh is gratefully acknowledged. We are thankful to Gabriel Pérez for his work on the sketch presented in Fig. \ref{fig:sketch}. This research has made use of data obtained from the \textit{Chandra} Data Archive and software (CIAO and TGcat) provided
by the \textit{Chandra} X-ray Center (CXC). 
\end{acknowledgements}

%
%

\bibliographystyle{aa}
\bibliography{Lib_paper}

\end{document}